\documentclass[amsmath,amssymb,nofootinbib,notitlepage,superscriptaddress,twocolumn]{revtex4-1}
\pdfoutput=1

\usepackage{aas_macros,graphicx}
\usepackage[bottom]{footmisc}
\usepackage[colorlinks=true]{hyperref}
\let\originalleft\left
\let\originalright\right
\renewcommand{\left}{\mathopen{}\mathclose\bgroup\originalleft}
\renewcommand{\right}{\aftergroup\egroup\originalright}

\input{epsf}
\usepackage[dvipsnames]{color}
\usepackage{txfonts}
\usepackage{euscript}

\usepackage{graphicx}
\usepackage{dcolumn}
\usepackage{bm}

\def\be{\begin{equation}}
\def\ee{\end{equation}}
\def\ba{\begin{eqnarray}}
\def\ea{\end{eqnarray}}

\def\ltsima{$\; \buildrel < \over \sim \;$}
\def\simlt{\lower.5ex\hbox{\ltsima}}
\def\gtsima{$\; \buildrel > \over \sim \;$}
\def\simgt{\lower.5ex\hbox{\gtsima}}

\definecolor{falured}{rgb}{0.5, 0.09, 0.09}

\begin{document}

\title{Flares and their echoes can help distinguish photon rings from black holes with space-Earth very long baseline interferometry}

\author{A. Andrianov}
\affiliation{Astro Space Center, Lebedev Physical Institute\thanks{Corresponding author A. Andrianov: andrian@asc.rssi.ru},
              84/32 Profsoyuznaya st., 117997 Moscow} 

\author{S. Chernov} 
\affiliation{Astro Space Center, Lebedev Physical Institute\thanks{Corresponding author A. Andrianov: andrian@asc.rssi.ru},
              84/32 Profsoyuznaya st., 117997 Moscow}

\author{I. Girin}
\affiliation{Astro Space Center, Lebedev Physical Institute\thanks{Corresponding author A. Andrianov: andrian@asc.rssi.ru},
              84/32 Profsoyuznaya st., 117997 Moscow}

\author{S. Likhachev} 
\affiliation{Astro Space Center, Lebedev Physical Institute\thanks{Corresponding author A. Andrianov: andrian@asc.rssi.ru},
              84/32 Profsoyuznaya st., 117997 Moscow}
\author{A. Lyakhovets} 
\affiliation{Astro Space Center, Lebedev Physical Institute\thanks{Corresponding author A. Andrianov: andrian@asc.rssi.ru},
              84/32 Profsoyuznaya st., 117997 Moscow}
\author{Yu. Shchekinov}            
\affiliation{Astro Space Center, Lebedev Physical Institute\thanks{Corresponding author A. Andrianov: andrian@asc.rssi.ru},
              84/32 Profsoyuznaya st., 117997 Moscow,\\
              and Raman Research Institute, Sadashivanagar, Bangalore}  
                           
\date{\today}

\begin{abstract}
Photon rings near the edge of a black hole shadow is supposed to be a unique tool to validate general relativity and provide reliable measurements of principal black hole parameters: spin and mass. Such measurements are possible though only for nearby supermassive black holes (SMBH) with Space-Earth Very Long Baseline Interferometry (S-VLBI) in the submillimeter wavelength range. For subrings to be distinguished  S-EVLBI observations with long baselines at the Lagrangian Sun-Earth L2 libration point are needed.  However, the average fluxes of nearby SMBH: Sagittarius A$^\ast$ (Sgr A$^\ast$) and M87$^\ast$ -- $F_\nu\sim 1$ Jy, are still insufficient to detect the signal from the photon rings with even such long baselines. We argue that only manifestations of flares in the submillimeter waveband in their accretion disks can reveal observable signals from the photon rings with the S-EVLBI at L2. Such observations will become possible within the planned join program of the {\it Event Horizon Telescope} (EHT) and {\it Millimetron Space Observatory} (MSO), and within the planned {\it next generation EHT} (ngEHT) project. Two different observational tests for photons rings are discussed. The first one involves observations of a time series of responds from subsequent subrings as can be seen in a 1D visibility function within the join EHT-MSO configuration, the second one -- measurements of an increase of the angle between subsequent subrings in the 2D VLBI image which can be obtained within the ngEHT project. 
\end{abstract}


\maketitle

\section{Introduction\label{intro}}
The Event Horizon Telescope Collaboration (EHTC) has approached the very vicinity of 
the supermassive black hole (SMBH) -- its shadow, in the center of galaxy M87 with 
the highest angular resolution of $\sim 25~\mu$as available within groud-based 
Very-Long-Baselines (VLBI) technique 
\cite{2019ApJ...875L...1E,2019ApJ...875L...2E,2019ApJ...875L...3E,2019ApJ...875L...4E,2019ApJ...875L...5E,2019ApJ...875L...6E}. This opens a new era of studying spacetime geometry and plasma processes under very strong gravitational forces in the area neighboring the SMBH horizon. The primary goal of the EHT experiment was to directly observe the black hole ``image'', i.e. the shadow around the black hole, and to see whether the spacetime geometry is described by the Kerr metric as follows from Einstein general relativity and satisfies as such the ``no-hair theorem'' \cite[see discussion in ][]{falcke13}. In April 2017, a critical experiment at $\lambda=1.3$ mm revealed the shadow of the angular size $\theta_{\rm shadow}\simeq 45~\mu$as as predicted \cite{falcke00a}, and led the EHTC to 
conclude that a metric of a supermassive black hole in the center of galaxy M87 
is consistent with the Kerr metric \cite{2019ApJ...875L...6E}, and with the mass close 
to the one inferred early from stellar kinematics in the central zone of galaxy M87 
\cite{Gebhardt2011}.

A complementary approach to test general relativity has been recently suggested by 
\cite{johnson20}. When passing around the black hole close
to the innermost stable circular orbit (ISCO) photons from accreting plasma are captured onto a quasi-stable orbit 
\cite{bardeen73,luminet79} where they can round over the ISCO a few times. Those 
reaching the observer screen leave on it a bright photon ring consisting of weaker 
embedded subrings, that imprint characteristics of spacetime metrics in which they have 
been formed. Thus they represent an informative tool for a quantitative characterization
of the metric, and provide an authentic validation of Einstein general relativity 
\cite{johnson20}. Observationally this information is encoded in a 1D visibility function
consisting of a hierarchy of responses from subsequent photon subrings. Each ring is formed by 
light that was emitted by the accretion disk and reached the observer's screen along $n$ half-orbits trajectories around the black hole. Those photons that come directly from the accretion disk are assigned $n=0$. The photons lensed along one half-orbit trajectory are focused onto the {\it leading} ring with $n=1$. As stressed by Ref.~\cite{johnson20} higher order subrings with $n>1$ unambiguously encode parameters of the black hole metric and can serve a unique tool for studying gravity  theories. The higher the order of a subring $n$ the slower the amplitude of its 1D visibility function declines on larger VLBI base lengths. For the visibility functions of $n=2$ and $n=3$ subrings to exceed contributions from the disk $n=0$ and the leading ring $n=1$, the baseline has to be increased beyond $u(\lambda)\simgt 3\times 10^{11}~\lambda$ (see Fig. 5 in Ref.~\cite{johnson20}). Therefore, observations of the rings $n=2$ and $n=3$ in the submillimeter waverange suggest baselines equivalent to the distance between Earth and the Lagrangian libration point L2.
However, with the central flux density of the order $\sim 1$ Jy in the nearby 
SMBH -- Sgr A$^\ast$ and M87$^\ast$, and consequently photon subrings are to have only
$\sim 100~\mu$Jy ($1\sigma$ level) at baselines $\sim 1~{\rm T}\lambda$ at the frequency 
$\nu=230$ GHz in the L2 point, [see Fig. \ref{Fig_1DVis} below in Sec. \ref{toy_1}]. 
Such a flux seems to be insufficient for detection with the planning MSO-EHT 
configuration at $\nu=230$ GHz with a bandwidth of 16 GHz and the ultimate $1\sigma$ 
detection limit $\sim 0.2$ mJy (see discussion in Ref.~\cite{Novikov2021}). 

A possible solution of this issue can be sought in utilizing enhanced brightness
of submillimeter (submm) flares on the accretion disk, via extracting similarities in time 
series in the main flare and its reflections on the photon subrings. Submillimeter flares
are known to accompany bright X-ray and near-infrared flares with time delays of $\sim 20$ 
minutes \cite{Witzel2021} to a few hours (see review and references in Ref.~\cite{Genzel2010}).
A month-long delay between the NIR and submm flares has occured recently (June 2019), as
described by \cite{Murchikova2021}. From comparison of its structure function  with one of
other similar flares taken place during the period 2002--2017, Ref.~\cite{Murchikova2021}
concluded that the most likely source of these flarings is magnetic reconnection. Numerical 
simulations performed by several research groups indicate that magnetic reconnections preferentially
occur in the innermost regions of accretion disks and even close to jet boundary 
\cite{Ripperda2020,Chatterjee2021}. This circumstance makes the submm wavebands a unique 
instrument for studying space-time metric near black holes, along with the properties
of relativistic plasma in extreme conditions close to the inner boundary of accretions disks 
and near the jet core.  In the current paper we analyze this possibility. The paper is 
organized as follows. Section \ref{disc_model} describes a simplified model of geometrically
thin disk and the black hole shadow. 
Section \ref{angul} considers the effects of the black hole spin on the photon rings that 
appear in our simplified model in the relationship between the parameters $a$, the spin, and 
$\theta_{\rm los}$, its inclination to the line of sight. Section \ref{time} describes a 
simplified model of a flare in the accretion disk and analyzes its interferometric response in form 
of a time-dependent visibility function on a long MSO-EHT baseline. In addition, we analyze a 
feasibility of VLBI imaging of a flare with its echoes in photon rings for Sgr A* with the ngEHT 
ground-based network. Section \ref{sum} contains conclusions. 


\section{Black hole shadow model}\label{disc_model}

\begin{figure}
   \centering
   \includegraphics[angle=-00,width=8cm]{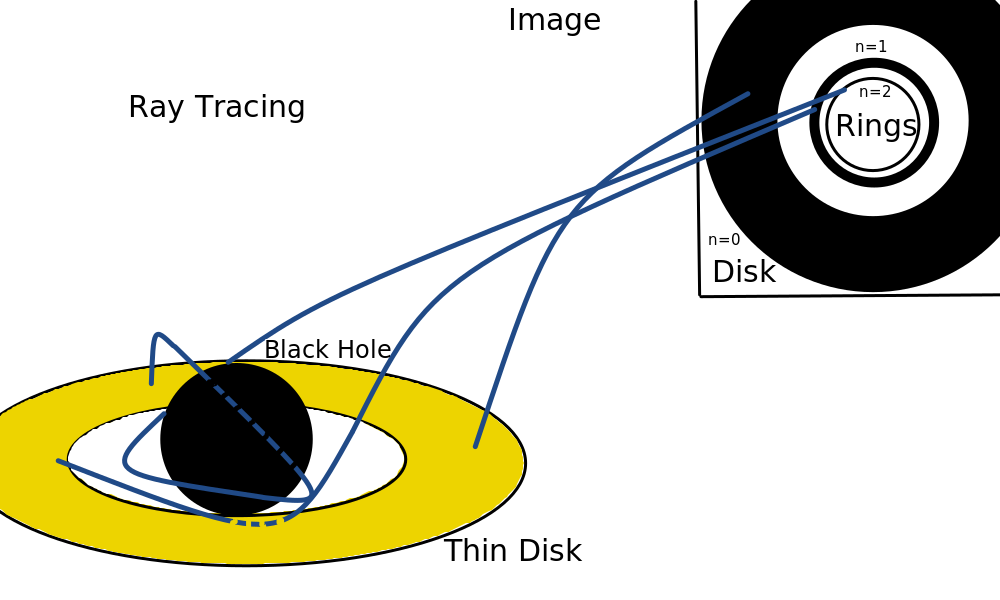}
      \caption{A simplified toy model (not to scale) of a geometrically thin disk around a Kerr
      black hole and the image of photon rings in the observer plane. The inner disk radius 
      is set to $r_{in}=8r_g$,  $r_g=GM/c^2$, the outer disk radius is $r_{out}=45r_g$. The resulting
      image on the observer's screen includes an image of the accretion disk (n=0) and images
      of the photon rings (n=1,2,3).
              }
         \label{Fig_model}
   \end{figure}

In this paper we use a simplified toy model of geometrically thin disk around
a Kerr black hole with the mass $M$ and angular momentum $J=aM$ ($0\leq a\leq M$) 
(see Fig. \ref{Fig_model}). The inner disk radius is set to $r_{in}=8r_g$,  $r_g=GM/c^2$, 
the outer disk radius is $r_{out}=45r_g$. In this simplified toy model we assume that 
the disk radiates uniformly (each pixel has equal brightness) and isotropically. 
The spacetime metric as described in Kerr-Schild
coordinates with $G=c=1$ is\cite{debney69,kerr65,balasin94,krasinski09}

     \begin{eqnarray}\label{metr}
      g_{\mu\nu}&=&\eta_{\mu\nu} + f k_\mu k_\nu \\
      f&=&\frac{2Mr}{\Sigma},~\Sigma=\frac{r^4+a^2 z^2}{r^2}\nonumber \\
      k_\mu&=&(1,k_x,k_y,k_z) \nonumber \\
      &=&\left(1,\frac{rx+ay}{r^2+a^2},\frac{ry-ax}{r^2+a^2},\frac{z}{r}\right)\nonumber
   \end{eqnarray}
where $k^\mu$ is the wave vector, $r$ is defined by~

\begin{equation}
\frac{x^2+y^2}{r^2+a^2}+\frac{z^2}{r^2} =1.
\end{equation}

The ``null'' geodesics for photon trajectories are decsribed by

    \begin{equation}\label{null}
    g_{\mu\nu}\frac{dx^\mu}{d\lambda}\frac{dx^\nu}{d\lambda}=0,
    \end{equation}
with $g_{\mu\nu}$ from Eq. (\ref{metr}), and $\mu,~\nu=0,~1,~2,~3$, $\lambda$ being affine parameter.

For numerical integration we can write Eq. (\ref{null}) as a system of eight coupled first-order equations:

    \begin{eqnarray}\label{metr_eq}
      \frac{d x^{\alpha}}{d \lambda} & = & k^{\alpha} \\
      \frac{d k^{\alpha}}{d \lambda} & = & - \Gamma^{\alpha}_{\mu \nu} k^{\mu} k^{\nu} \nonumber
    \end{eqnarray}
where $\Gamma^{\alpha}_{\mu \nu}$ are the Christoffel symbols.

Let us consider a distant observer whose line-of-sight inclination angle with respect to the 
black-hole rotation axis is $\theta_{\rm los}$. The vertical axis of the observer's image plane
$\beta$ is directed toward the projection of the black hole spin provided $\theta_{\rm los}\neq 0$.
On this plane, $\alpha$ is the impact parameter perpendicular to  $\beta$ \cite{Cunnigham1973}. 
For obtaining image on the observer's screen we apply the backward ray tracing for each image
pixel ($\alpha, \beta$). The initial wave vector $k$ is directed to the center of the black hole --
the point with coordinates (0,0,0). In numerical calculations the distance of the observer's 
screen from the BH is set to 40$r_g$. For each image pixel, Eqs. \ref{metr_eq} are solved with
the Runge-Kutta 4-order integration method.

   \section{Angular displacements of the rings}\label{angul}

Figure \ref{Fig_model} illustrates possible photon trajectories that form the photon rings in our model. As mentioned in Sec. \ref{intro} the photons coming directly from the disk form its image on the observer's screen, those bended by lensing onto the one-half orbit trajectory $n=1$ are focused into the leading ring, whereas $n>1$ photons form higher order subrings. 

Fluxes transmitted to subsequent photon rings depend on the accretion flow geometry, 
radiating plasma properties, and the path length through the emitting plasma \cite{Gralla2019,johnson20}.
General relativity dictates that relative contributions of subsequent rings are predominantly 
determined by the BH spin and its inclination to the observer. As the photon rings collect photons 
from wide areas in the disk they are brighter for higher $n$, however the total flux in a given 
ring decreases with $n$. For a Schwarzschild BH, the next order photon ring $n+1$ carries only 4\% of
the total flux in the previous $n$ ring   \cite{luminet79,Gralla2019}. For $a/M\sim 1$ and the
inclination angle $\theta_{\rm los}$ of $\sim 17 ^\circ$ it can be of $\sim 13$\% \cite{johnson20}.
In our simplified model with the geometrically thin uniform disk this parameter is $\sim 10$\% . 

   \begin{figure}
   \centering
   \includegraphics[angle=-00,width=8cm]{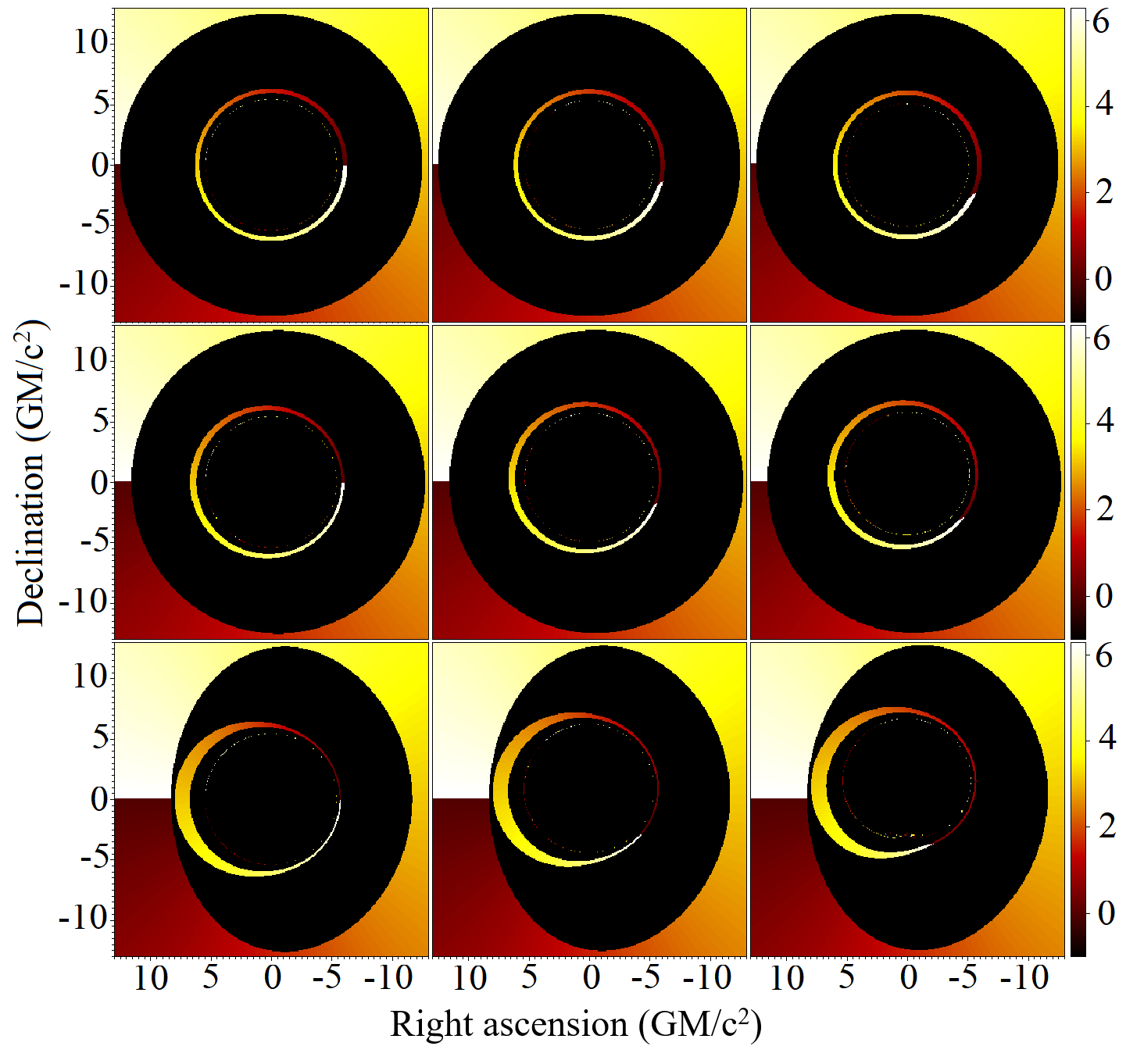}
      \caption{Photon rings from a thin disk are color coded white to brown depending on the angle
      from a given direction {(coincident with $\beta$-axis -- the spin projection 
      onto the observer's screen)}. The disk projection on the leading photon ring is similarly coded.
      {\it The upper row} shows the disk and the leading photon ring for a black hole with 
      angular momentum along the sight line $\theta_{\rm los}=0$, $a=0.0,~0.5,~1.0$ from left 
      to right; {\it the middle row} shows the same but for angular momentum inclined to the sight 
      line by $\theta_{\rm los}=17^\circ$, and {\it the lower row} is for $\theta_{\rm los}=45^\circ$.
      The turn of the disk pattern map reflected on the photon ring is clearly seen from {\it i)}
      increase in $a$ shown on panels from left to right, and {\it ii)} increase in the spin 
      inclination to the line of sight $\theta_{\rm los}$ -- panels from top to bottom.
              }
         \label{Fig_turn}
   \end{figure}

Figure \ref{Fig_turn} shows the modelled angular brightness distribution on the disk 
$I(\phi)$ and its projection onto the image plane. We color the disk white to brown
clockwise starting from the direction coincident with the spin projection in the image plane as shown in Fig. \ref{Fig_turn}. Where $\theta_{\rm los}=0$ and $a=0$, as in the upper left subimage, the edge between brown and white is fixed along $\alpha$-axis. In this case ($\theta_{\rm los}=0$, $a=0$), the on-disk color pattern mapped onto the photon ring turns clockwise by $\delta\phi_{obs}=180^\circ$ \cite{luminet79,Gralla2020}, again from white to brown.  The panels in the upper row, left to right, correspond to the spin increase from $a=0$ to $a=1.0$ with increment $\delta a=0.5$, the color pattern turns accordingly by $\delta\phi\approx 25^\circ\delta a$, in addition to  $\delta\phi_{obs}=180^\circ$. As the inclination $\theta_{\rm los}$ increases the color pattern angle $\phi$ grows $\delta\phi(a,\theta_{\rm los})$, as can be derived by solving numerically Eq. \ref{metr_eq}, or analytically from Eqs. (7b), (8c), and (8d) in \cite{Gralla2020}. Detection of bright sporadic flares (and their echoes in the photon rings) on the VLBI images can be utilized to distinguish photon rings and reveal the angular correspondence between events on the disk and in rings. For Sgr A*, this possibility is shown in Section \ref{toy_sgr}.

   \section{Time variability in the disk}\label{time}

\begin{figure}
   \centering
   \includegraphics[angle=-00,width=8cm]{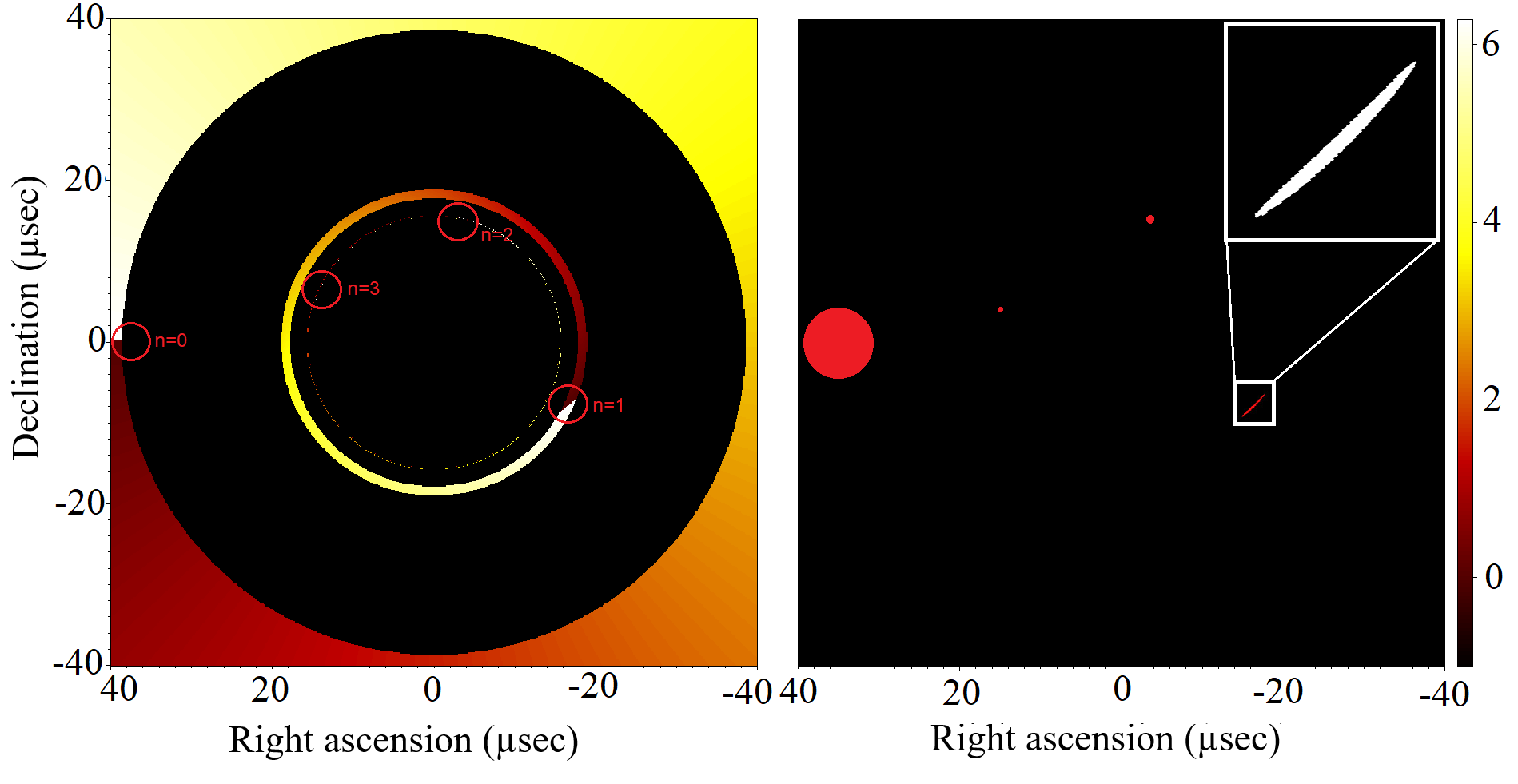}
      \caption{{\it On the left panel} small red circles show: location of 
      a flare spot in a selected region {at the boundary between the bright and dim areas}
      of the disk, and location of its reflections on the rings $n=1,~2$ and 3. 
      The brightest and widest ring is the leading one with $n=1$, the subring with $n=2$
      is seen as a very thin ring inward, the $n=3$ subring further inward is not resolved in
      this figure. As in Fig. \ref{Fig_turn}, the colors code the anglular position on the disk. 
      {\it On the right panel} we show the flare position on the accretion disk as 
      a uniform red spot. It reflects on photon rings as spots of progressively smaller
      size, as seen on the right panel. The small filled circles on the right show the images of
      reflections on the observer's plane. The $n=1$ reflection {is shown in its real shape.
      Reflections
      with $n=2$ and $n=3$ are plotted schematically as small red points at their positions.} 
      {In the upper right corner the shape of $n=1$ reflection is shown at a larger scale.}
      The total brightness of each $n$-th order subring is $\approx 0.1$ of the previous
      $(n-1)$-th one \cite{johnson20}. 
              }
         \label{Fig1sub_reverb}
   \end{figure}

Bright sporadic flares (and their sequences) can be utilized to distinguish photon
rings from the main crescent image of the disk. Let us assume a spot-like flare at
time point $t_f$ and at a given location in the disk. On the observer's plane the flare first
responds in the crescent image, and {after time delay $\Delta t_f\sim {\rm a~few~}\times n\pi r_{\rm g}/c$ 
repeats its respond in the projection onto the leading photon ring ($n=1$), and subsequently
onto subrings with higher $n$}. Overall, each of such flares can be reflected onto subsequent 
photon rings with a sequence of subflares as illustrated in Fig. \ref{Fig1sub_reverb}.
The geometric shapes of a flare and its projections onto the first ring are shown
as red spots in the right panel of Fig. \ref{Fig1sub_reverb} and have been 
numerically obtained by  backward ray tracing of the geometrically thin accretion disk 
model (See Section \ref{disc_model}). Note that the flare responds on the rings are 
brighter because they are localized in a much narrower area on the rings contrary to its 
spread respond on the crescent. Submm flares are apparently very compact as compared 
to the accretion disk. On the long space-ground baseline, however, they produce 
fluxes comparable to the average flux from the disk (see discussion below). This
circumstance can facilitate better observatility of the photon ring and subrings on a short
time scale.  A very recent study Ref.~\cite{chesler2021} explores in detail
how such reverberation of sporadic brightness 
variability in the accretion disk on photon rings manifests in single-dish
observations. 

   \begin{figure}
   \centering
   \includegraphics[angle=-00,width=8cm]{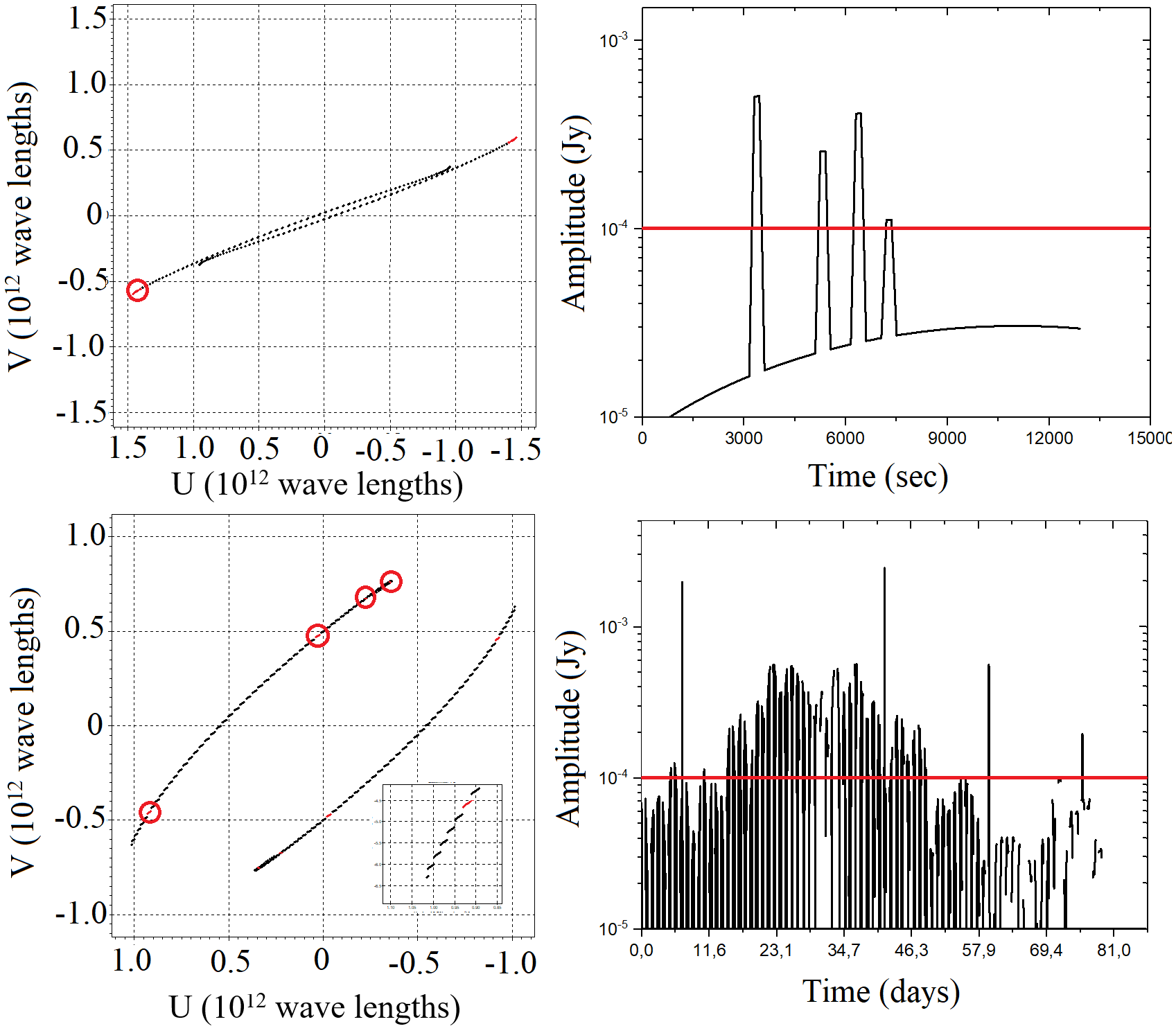}
      \caption{{\it The left panels} show a typical $uv$ coverage that will be available with the ALMA-MSO interferometer in the orbit around the L2 point. {\it The right panels} show the time dependence of the visibility amplitude for short flares as if they occurred in the accretion disk  as shown schematically in Fig. \ref{Fig1sub_reverb}. The upper row shows the case of Sgr A*, the lower row shows the case of M87*. The visibility function has been synthetically measured in the time range of approximately 60 full days for M87 and approximately 1 hour for Sgr A*. Open circles on the UV-tracks on the left panels indicate the times corresponding to the flare and its echoes. The total disk flux (i.e. the flux from $n = 0$-th order ring) is normalized to 1 Jy, flare magnitude in the disk is set to $2$ Jy. The thick horizontal line in the right panels shows the 0.1 mJy sensitivity level ($1\sigma$) planned for the ALMA-MSO baseline.  
              }
         \label{Fig_MM_UV}
   \end{figure}

Flares of brightness in the NIR and submm wavelength domains are observed in Sgr A$^\ast$. Their duration varies from a few minutes, with a median delay of $\sim 20$ min in the submm to NIR \cite{Dodds2009,Do2019,Witzel2021}, to $\sim 1-4$ hours \cite{Mauerhan2005,Marrone2006,Marrone2008,Eckart2008,Trap2011}. In millimeter (mm) and submm wavebands, the relative magnitude of variability of Sgr A$^\ast$, i.e. the ratio of the observed peak to the minimum flux $F_{\rm max}/F_{\rm min}$, can reach the order of $\sim 2$  \cite[][]{Zhao2003,Trap2011,Dexter2014,Witzel2021,Michail21}. At higher frequencies this ratio increases approximately as $F_{\rm max}/F_{\rm min}\propto \nu^{1/2}$ as inferred in Ref.~\cite{Genzel2010}, see their Fig. 31.  

The origin of these flares is still unclear, and possible mechanisms are widely debated (see discussions in Ref.~\cite{Garcia2011}). Recently discussed related mechanisms include flarings from magnetic reconnection with associated flux eruptions \cite{Dexter2020,Petersen2020,Porth2020,Gutierrez2020} in the regime of a magnetically arrested accretion disk (MAD) in a near-horizon domain, $r\sim (10-40)r_g$ \cite{Ripperda2020,Ripperda2021,Chashkina2021}. Lack of powerful reconnections in a standard accretion and normal evolution (SANE) model makes it inefficient in producing bright flares at high frequencies $\nu>10^{14}$ Hz, as recently confirmed by Ref.~\cite{Scepi2021}. Within the concept of MAD flows, the characteristic time and scale are determined by the reconnection rate and the time of a few $r_g/c$ needed for re-establishing the quasi-steady-state accretion after the flare \cite{Ripperda2021}, in accordance with polarimetric measurements of the horizon scale $r\sim (6-10)r_g$ magnetic field in Sgr A$^\ast$ by the GRAVITY observations \cite{Gravity2018}.

{In the case of Sgr A$^\ast$, stellar winds from Wolf-Rayet stars within the central 
parsec (from $\sim 1$ pc in the outer disk toward $\sim 300~r_g$ in the inner domain 
of the Galactic Centre) can be an important source of maintaining MAD accretion 
\cite{Cuadra2008,Cuadra2015,Resler2018,Ressler2020a,Ressler2020b,Calderon2020}. 
Ref.~\cite{Ressler2019} presented a model of a stellar wind-fed MAD in Sgr A$^\ast$ that 
matches observational data and the time variability of rotation and 
dispersion measures, among other things (see Ref.~\cite{Ressler2020a} for more discussion and references). 
Previous estimates and conclusions  that the flares are compact with sizes much less than
the accretion disk, are largely consistent with these predictions.}  Ref.~\cite{Doeleman2008}
put an upper limit for their sizes $\simlt 4r_g$. Ref.~\cite{Witzel2018} argued that a 
reasonable estimate from time delay between submm, near-infrared and X-ray can be $\approx 2 r_g$.
Recently Ref.~\cite{Michail21} scrutinized a submillimeter flare in Sgr A$^\ast$, and 
concluded from time-delay between submillimeter and infrared that it is connected with a plasma
blob of size $\sim r_g$. Even longer submm flares with $\Delta t\sim 1$ hour can be confined 
into compact regions of order $\sim r_g$ if the emitting plasma blobs expand adiabatically at 
a speed of $v\sim 0.005c$, as shown by Ref.~\cite{Eckart2008}.
More recent GRAVITY observations \cite{Gravity2018} spatially resolved three flaring blobs
with orbital motions within $r\sim (6-10)r_g$ \cite{Ball2021,Porth2020}, i.e. close to the black hole.
It is worth 
mentioning that even though the flares originate in the current sheet very close to or even partly
within the ISCO region in the domain of jet formation, some of plasmoid blobs and associated 
flares still remain on stable orbits around the disk, as demonstrated by the GRAVITY experiment 
\cite{Gravity2018}, and as such are not channeled into the jet.

  \subsection{Interferometer visibility function for a single flare}\label{toy_1}

   \begin{figure}
   \centering
   \includegraphics[angle=-00,width=8cm]{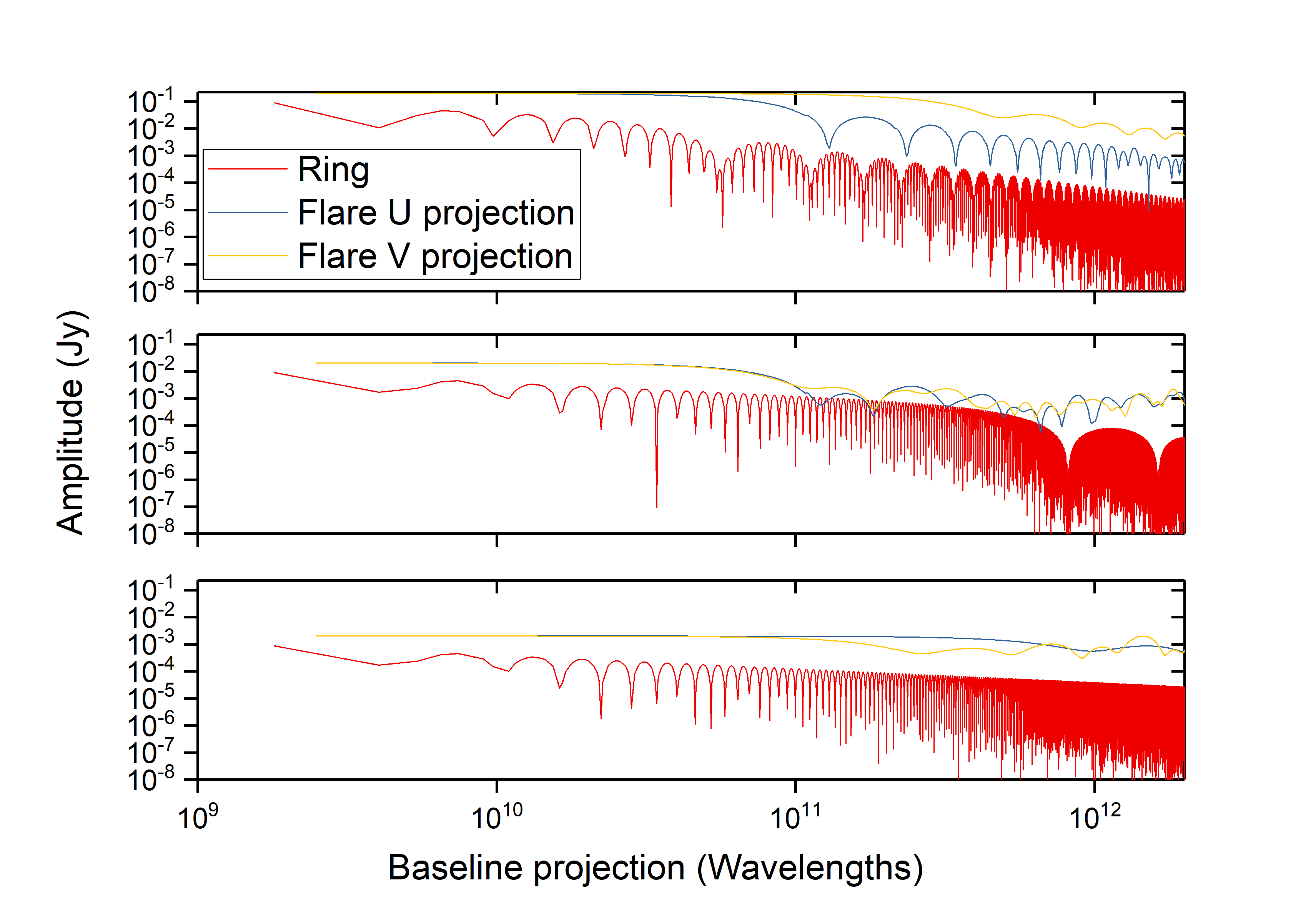}
      \caption{1D visibility functions $V$ along $u$ and $v$, panels from top to bottom show responds of different orders $n=1,~2,~3$ from the photon rings themselves (lower curves) and the flares (upper curves): the darker (blue) lines present $u$-projections, whereas the pale (yellow) ones correspond to $v$-projections). The baseline is given in wavelengths, the total flux from the ring $n=0$ (accretion disk) is normalized to 1 Jy; that from the flare in the disk $n=0$ is normalized to 2 Jy; based on Ref.~\cite{johnson20} we assume the total flux for each $n>0$ to be 10\% of the previous one $n-1$.}
         \label{Fig_1DVis}
   \end{figure}

Asymptotically, the partial contribution of a photon ring with a given $n$ into the visibility function at a baseline longer than $u\gg w_n^{-1}$ decays as $\propto u^{-1/2}e^{(-uw_n)^\zeta}$.
Here, $w_n$ is the width of $n$-th ring,
and $\zeta$ is determined by the radial profile of the photon ring \cite{johnson20}.
Evidently, photon rings with smaller $n$ and hence larger $w_n$ decay faster.
{In our model, the visibility function on longer baselines for a uniform ring of a finite width $w_n$ is determined directly as \cite{Chernov2021}
\be
V_n(u)={1\over \pi d_n w_nu}\left(b_nJ_1(2\pi b_nu)-a_nJ_1(2\pi a_nu)\right),
\ee
where $a_n$ is the inner, and $b_n$ the outer radius of $n$-th ring,  $d_n=a_n+b_n$, and $b_n-a_n=w_n$, the width of the ring}. As a consequence, on longer baselines the visibility amplitudes $V_n$ with higher $n$ become distinguishable from the amplitudes of the leading one, $V_1$: at  $u \sim {10}~{\rm G}\lambda$, the subring with $n=2$ dominates, while at $u > {1000}~{\rm G}\lambda$, the subring $n=3$ comes into play, as seen in Fig. \ref{Fig_1DVis}. It is clearly seen though that the amplitude of the visibility function at long baselines declines dramatically and can fall below the sensitivity threshold of the instrument under use. Indeed, for M87$^\ast$, with the central flux density of $\sim 1$ Jy, photon rings with $n=3$ asymptotically decay by at least a factor of 30, and hence they cannot be resolved by the currently used or planned instruments, as seen in Fig. \ref{Fig_1DVis}. 

An alternative can be found in S-EVLBI observations of sufficiently bright flares in the accretion disk. {Fig.} \ref{Fig_1DVis} also shows projections on U and V of the visibility function corresponding to the reflection of a single flare in the accretion disk on the corresponding photon ring. {The flare compactness results in very compact echoes on rings, as seen in the ray-tracing shape of the $n=1$ reflection in the upper right corner of Fig. \ref{Fig1sub_reverb}. Therefore, despite the fact that the total flare flux is only twice the total flux from the entire quiet disk, its response seen in the amplitude of the visibility function on long interferometer baselines can be more than an order of magnitude higher than the visibility function of the corresponding ring.} Thus, observation of VLBI flare signatures with long space-ground baselines requires an order of magnitude softer sensitivity requirements than {those for observation of VLBI photon ring signatures from a smooth quiet disk.}

  \subsection{Time variability of interferometer visibility function for a flare}\label{toy_2}

Let us assume that a spatially localized short-term flare of $\Delta t_f<5\pi r_{\rm g}/c$ has occurred on the disk (see left panel on Fig. \ref{Fig1sub_reverb}). The flare image reflects subsequently onto photon rings (Fig. \ref{Fig1sub_reverb}), with the time gap between the subsequent subflares $\Delta t_g$ nearly equal to $\sim 5\pi r_{\rm g}/c$. The flare on the disk will first manifest in the visibility function in the central domain of $u$, and afterwards, with a time delay $\delta t_d\sim \Delta t_g$, it will subsequently appear in the peripheral domains of $u$ corresponding to photon rings with an increasing $n$. The exact value of $\delta t_d$ {was calculated from the model using ray tracing}. The advantage related to such manifestations is that each subflare on the corresponding subring does appear separately of the previous brighter subflare, and thus can be distinguished easier. An example of how a flare on the disk will reverberate on the visibility function with the ALMA-MSO S-VLBI is demonstrated in Fig. \ref{Fig_MM_UV}. The left panels show a typical $uv$ coverage expected within  ALMA-MSO interferometric observations of Sgr A* (top row) and M87* (bottom row)  in orbit around the L2 point. The right panels show the time dependence of the measured visibility amplitude for short flares. For the flares, we assume the magnitude of twice the steady state accretion disk \cite{Witzel2021}, i.e. 2 Jy for Sgr A$^*$. {Figure \ref{Fig_1DVis} depicts 1D visibility functions for the flare schematically shown in Figure \ref{Fig1sub_reverb} and subsequently reflected on subrings $n=1,~2,~3$ (top to bottom). Lower curves in the panels show visibility functions $V(u)$ of subrings from a quiet disk, the upper curves correspond to visibility functions $V(u)$ (blue) and $V(v)$ (yellow) in $u$ and $v$ projections, respectively. As seen, the amplitude of the visibility function of echoes from the flare on the long space-ground baseline is more than an order of magnitude higher than the one of the corresponding ring itself, because of the compactness of the flare compared to the disk.
This results in an order of magnitude weaker sensitivity requirements for the detection of VLBI flare signatures
with long space-ground baselines
than sensitivity needed to detect VLBI photon ring signatures from a smooth quiet disk. } 

  \begin{figure}
   \centering
   \includegraphics[angle=-00,width=8cm]{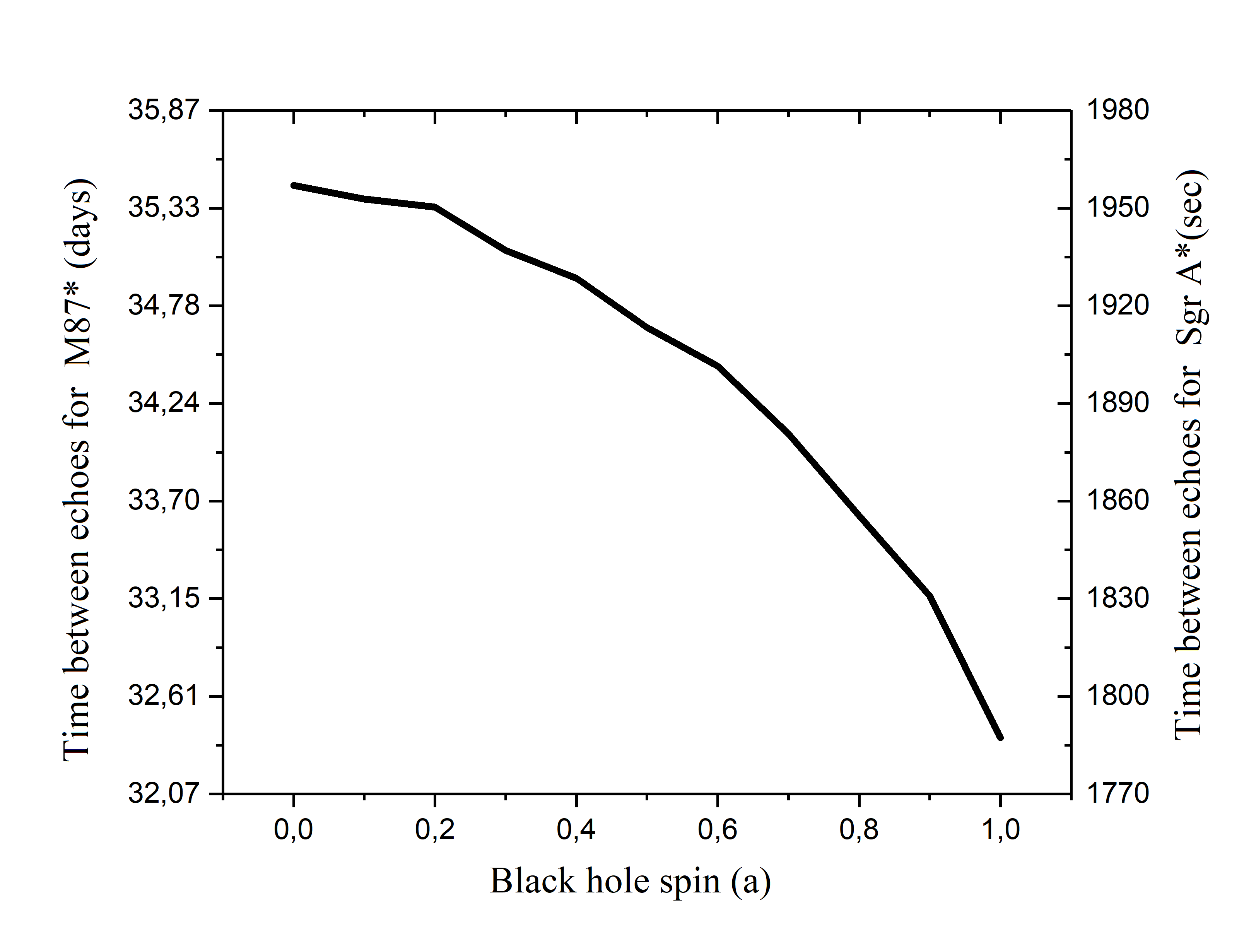}
      \caption{
The plot shows time delay between the flares in the third and the first photon rings $\Delta t$ versus the black hole spin for M87* (left $y$-axis) and Sgr A* (right $y$-axis), $\theta_{\rm los} = 0$. Relativity short  ($\simlt 1$ hour for Sgr A* and $\simlt 3$ days for M87*) flux monitoring is {sufficient for the flare echoes to be distinguished}.  
              }
         \label{Fig_deltaT_from_a}
   \end{figure}

The time lag between the arrival times of the echo from the flare in the first and subsequent rings can be used to determine the parameter $a$. Fig. \ref{Fig_deltaT_from_a} shows the time delay between the flare copies in the third and the first photon rings versus the black hole spin for M87* and Sgr A*. {Due to the mass difference the reverberation time lag between the echoes in the first and third ring in the case of M87* is considerably longer -- about 30 days as compared with the short lag $\simlt $ {several minutes}  in Sgr A$^\ast$, and seems sufficient for monitoring the reverberation on a known date  {(within $\simlt 3$ days window for M87*)} with the S-E VLBI baseline in both cases. For instance, in the case of Sgr A*, flux monitoring for about 2 hours can bring a pronounced conclusive result.}

   \begin{figure}
   \centering
   \includegraphics[angle=-00,width=8cm]{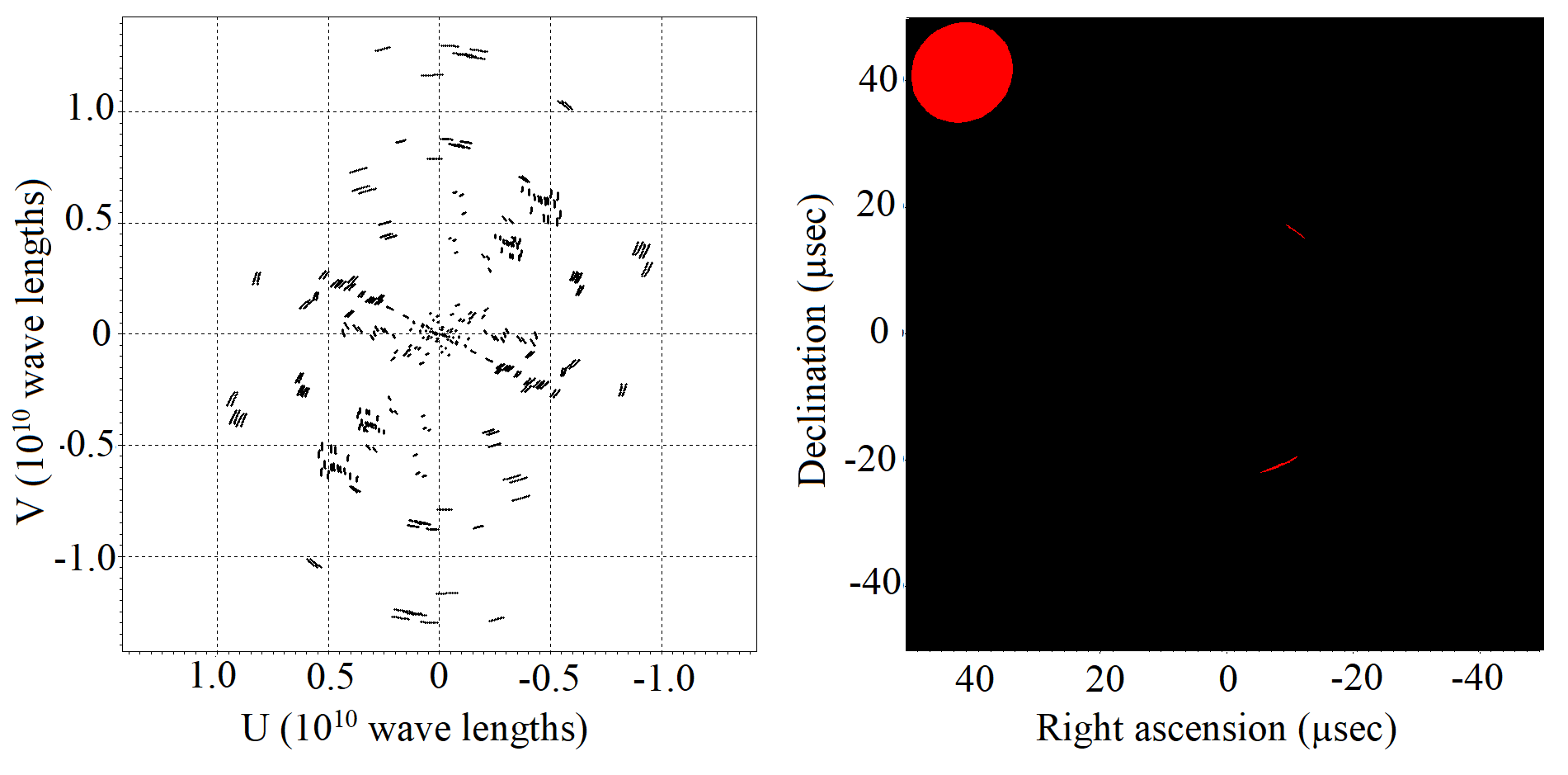}
      \caption{Left panel: UV-coverage for a synthetic instantaneous ngEHT observation of Sgr A*. Duration of observations is assumed 30 minutes, UV plane coverage is composed for two frequencies 230 and 340 GHz, the sites and parameters of the ngEHT array are taken from Ref.~\cite{Blackburn2019}. Right panel: VLBI model image for a long flare.  It consists of a flare in the thin accretion disk with a total flux of 2 Jy and two echoes in the photon rings. The total flux in the first echo is 0.2 Jy and 20 mJy in the second one, the black hole spin is taken $a = 0.8$, the inclination angle $\theta_{\rm los} = 0$. Model image is obtained by ray tracing.
              }
         \label{Fig_SGR_1}
   \end{figure}

   \begin{figure}
   \centering
   \includegraphics[angle=-00,width=8cm]{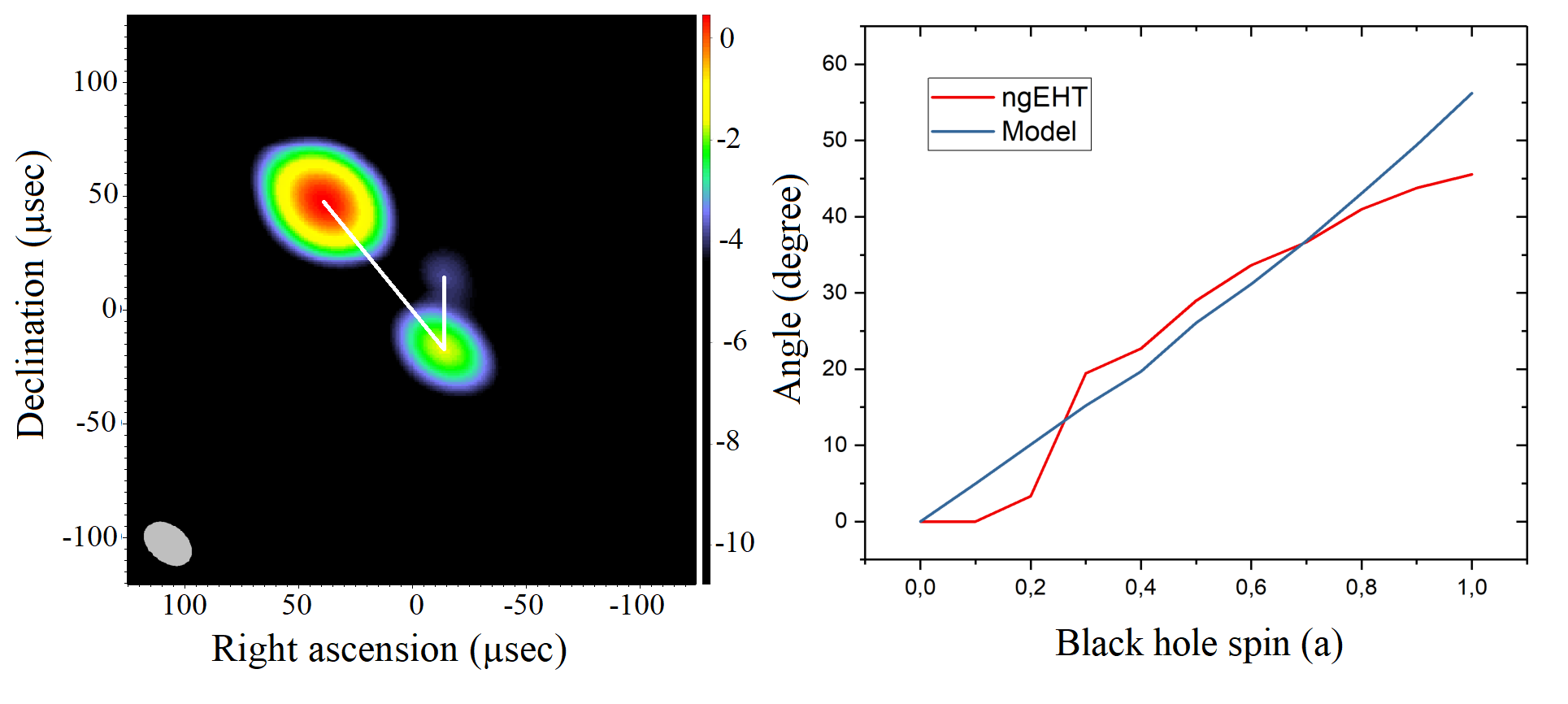}
      \caption{ Left panel: reconstructed instantaneous VLBI image of a {long} flare in the accretion disk and two echoes in the photon ring with the ngEHT VLBI array by CLEAN method. The lower left corner shows the beam size, the array parameters are taken from Ref. \cite{Blackburn2019}, UV-coverage and source model are plotted at Fig \ref{Fig_SGR_1}. White lines mark the angle $\alpha$ between the direction of the flare in the accretion disk and the direction of its echo in the {$n=1$ photon ring}, the intensity is plotted in logarithmic scale {(Jy/beam)}. Right panel: dependence between the angle $\alpha$ and the black hole spin $a$. The positions of the spots on the VLBI image are determined from the intensity maximum. The broken line corresponds to the restored ngEHT VLBI image, the straight one corresponds to the model. As seen, the angular resolution and 30-minutes UV-coverage of the ngEHT is sufficient to measure the angle $\alpha$.
              }
         \label{Fig_SGR_2}
   \end{figure}

  \subsection{Angular displacements of the VLBI flare image for Sgr A*}\label{toy_sgr}

Longer submm flares in Sgr A* with $\Delta t_f > 5\pi r_{\rm g}/c$ can be represented in the VLBI image as a superposition of the flare in the accretion disk and its echoes on the photon rings. {An example of a synthetic observation with the ngEHT array is shown on the right panel of Fig. \ref{Fig_SGR_1}. The left panel of Fig. \ref{Fig_SGR_1} shows the UV-coverage for a synthetic ``instantaneous'' ngEHT observation of Sgr A* for 30 minutes. The UV plane coverage is composed of two frequencies 230 and 340 GHz. The sites and parameters of the ngEHT array are taken from Ref.~\cite{Blackburn2019}.  The right panel of this figure presents the VLBI model image of a long flare.  The image shows a flare in the thin accretion disk with a total flux of 2 Jy \cite{Witzel2021} and two echoes in the photon rings. The total flux in the first echo is 0.2 Jy and in the second echo 20 mJy. Black hole spin a = 0.8,  $\theta_{\rm los} = 0$. The model image was obtained by ray tracing.} In Sgr A*, the characteristic time {of long flares} in millimeter wavelenghts is 30 min -- 2 hours \cite{Yusef2008}, at least ten times longer than the gravitation time $\sim 200$ s. Therefore, for Sgr A$^*$, several bright spots will be simultaneously visible in the VLBI image (see the right panel of Fig. \ref{Fig_SGR_1} and the left panel of Fig. \ref{Fig_SGR_2}). The relative positions of these bright spots can be used to directly determine the parameters $a$ and $\theta_{\rm los}$. {It is important to note} that the angular resolution of the ground VLBI is sufficient to accurately determine the coordinates of bright spots in the image, {provided that the UV-plane has a good filling and the array ensures good sensitivity, as planned in the ngEHT}. Fig. \ref{Fig_SGR_2} (left panel) shows the reconstructed instantaneous VLBI image by the CLEAN method. The white lines mark the angle $\alpha$ between the direction of the flare in the accretion disk and the direction of its echo in {the $n=1$ photon ring}. The intensity {(Jy/beam)} is plotted in logarithmic scale. The right panel shows the dependence of the angle $\alpha$ on the black hole spin parameter $a$. The positions of the spots in the VLBI image {were assumed to coincide with intensity maximums. The broken line corresponds to the restored ngEHT VLBI image, the straight line is from the model.}
The sensitivity provided by a key anchor station of the ngEHT jointly with a small 10-m dish
is about 5-20 mJy  (Fig. 7 in Ref.~ \cite{Blackburn2019}),
{which corresponds to 2 GHz bandwidth and a few seconds} atmospheric timescale.
The planned bandwidth in the ngEHT is 16 GHz, so it will improve the sensitivity { $\approx \sqrt{16~{\rm GHz}/2~{\rm GHz}} \approx 2.8$ times}. Simultaneous observations at two frequencies using the FPT technique \cite{Guang2018} will increase the currently short coherence time of a few seconds (due to atmosphere) to about 150 s (due to the H-maser stability on the site). Thus, theoretically the VLBI sensitivity can be increased  $\approx \sqrt{{\rm 150 s}/{\rm 3 s}} \approx 7$ times more, and one may conclude that the planned ngEHT sensitivity will be sufficient to detect bright spots in the accretion disk and their echoes on the first and the second photon rings.

   \section{Summary}\label{sum}

In this paper, we analyzed a feasibility to observe manifestations of photon rings around the shadow of a supermassive black hole and to obtain information about the spacetime metric. Within a simplified model we showed that 

\begin{enumerate} 
\item SE-VLBI observations with the joint {EHT-MSO program} seem promising to detect manifestations of photon rings $n=1,~2$ from bright (with a factor of 2 magnitude enhancement) flares, provided that the maximum visibility function $V(0)\approx 1$ Jy, as for M87$^*$ \cite{2019ApJ...875L...6E}. Generally, the signal can be distinguished on baselines achievable by the joint EHT-MSO observations with the anticipated {$1\sigma$} sensitivity $\delta S\sim 0.1\hbox{--}0.3$ mJy at the Lagrange libration point  L2 with $u\sim 1~{\rm T}\lambda$ \cite{Novikov2021}. Despite the fact that the total flare flux is only twice the total flux from the corresponding ring, on a long interferometer baseline the flare can give more than an order of magnitude greater amplitude of the visibility function due to the fact that the flare itself concentrates the emission flux comparable to that from the entire disk, and the echo from the flare on the ring is much more compact than the respective ring. Thus, observation of VLBI flare signatures with long space-ground baselines requires an order of magnitude softer sensitivity requirements than observation of VLBI photon ring signatures from a quiet disk.
\item The brightness distribution on photon rings reflects the disk brightness with a time delay. The short strong flare in M87* and Sgr A* observed on long EHT-MSO baseline produces a series of subflares on photon rings with definite time delays  between them. The relative times between the subflares provide a possibility to directly determine the parameters $a$ and $\theta_{\rm los}$ of the black hole. 
\item The brightness distribution on photon rings reflects the disk brightness with an angular displacement. As a result, a single flare on the disk can produce several bright spots on photon subrings with a short time delay $\Delta t_f$ in the VLBI image. From the relative positions of these spots, one can directly determine the parameters $a$ and $\theta_{\rm los}$ of the black hole. The angular resolution, the quality of 30-min UV-coverage and the sensitivity of the ground-based ngEHT are  sufficient to resolve at least three (one in the accretion disk and two in the subring reflections) of these bright spots in SgrA*.

\end{enumerate}

\begin{acknowledgements}
      SVC was supported by RFBR grant 20-02-00469. The work of YS is done under partial support from the project ``New Scientific Groups LPI'' 41-2020. 
\end{acknowledgements}

%
%


\bibliography{pr_aps}
\bibliographystyle{utphys}

\begin{appendix} 
\end{appendix} 

\end{document}